\documentclass[aps,prl,twocolumn,superscriptaddress, footinbib, logbibliography]{revtex4-1}
\usepackage{xcolor}

\usepackage{graphicx}
\begin{document}

%\preprint{}

\title{Super-stretchable Elastomer from Cross-linked Ring Polymers}


\author{Jiuling Wang}
%\thanks{}
\affiliation{Department of Chemistry and Biochemistry, University of South Carolina,
Columbia, South Carolina 29208, USA}
\author{Thomas C. O'Connor}
\affiliation{Department of Materials Science and Engineering, Carnegie Mellon University,	Pittsburg, PA 15213, USA}
\author{Gary S. Grest}
\affiliation{Sandia National Laboratories, Albuquerque, New Mexico 87185, USA}
\author{Ting Ge}
\email[]{tingg@mailbox.sc.edu}
\affiliation{Department of Chemistry and Biochemistry, University of South Carolina,
	Columbia, South Carolina 29208, USA}

%Collaboration name if desired (requires use of superscript address

\date{\today}

\begin{abstract}		
The stretchability of polymeric materials is critical to many applications such as stretchable electronics and soft robotics,  yet the stretchability of conventional cross-linked linear polymers is limited by the entanglements between polymer chains. We show using molecular dynamics simulations that cross-linked ring polymers are significantly more stretchable than cross-linked linear polymers.  Compared to linear polymers, the entanglements between ring polymers do not act as effective cross-links. As a result, the stretchability of cross-linked ring polymers is determined by the maximum extension of polymer strands between cross-links, rather than between entanglements as in cross-linked linear polymers.  The more compact conformation of ring polymers before deformation also contributes to the increase in stretchability. 
\end{abstract}

\maketitle

%\section{}

Cross-linked elastomers are widely used as both flexible soft materials and highly deformable matrices for polymer composites.  
Elastomers are distinguished from metallic, ceramic, and other amorphous materials by their ability to reversibly accommodate large stretches $\sim 100$--$500\%$.
This makes them ideal for applications in soft robotics \cite{Gul2018},  wearable electronics \cite{Amjadi2016, Wu2019}, and biomedical devices \cite{Yeo2016}. 
However, elastomer softness and stretchability are limited by the tendency of polymer chains to form entanglements which topologically hinder the motion of network strands. 
Most polymers naturally form a dense network of entanglements at equilibrium, which dominate both the stiffness and stretchability of an elastomer if the entanglement density is higher than the cross-link density \cite{Pearson1980}.  Achieving ultra soft and stretchable elastomers requires overcoming the barrier produced by polymer entanglements.

One promising route toward reducing the entanglement density and enhancing the stretchability of elastomers is by controlling polymer chain topology through advances in chemical synthesis \cite{Gu2020}.
This is exemplified by the super-soft elastomers recently made from cross-linked bottle-brush polymers \cite{Cai2015, Cao2015, Daniel2016, Sheiko2019},  in which the side chains expand the effective diameter of the network strands and reduce the number of trapped entanglements.  
In this Letter, we investigate an alternative scheme based on the unique properties of non-concatenated ring polymers.  The topological constraints of non-concatenation force the ring polymers to have loopy and globular conformations \cite{Obukhov2014, Gooben2014, Rosa2014, Ge2016, Kruteva2020a}.  However, these topological constraints evolve with time in a self-similar manner \cite{Ge2016, Kruteva2020} such that there is no apparent entanglement network that strongly confines the dynamics of chains at any specific scale.  As a result, non-concatenated ring polymers possess distinctive linear \cite{Kapnistos2008, Pasquino2013} and non-linear \cite{Huang2019, Connor2020, Parisi2021}  rheology.  As we will show, the absence of an entanglement network in ring polymer melts offers a novel pathway to create ultra soft and stretchable elastomers. 

We use molecular dynamics simulations to generate and study the mechanics of ring polymer elastomers with precisely controlled topology. We show that non-concatenated ring elastomers exhibit dramatically higher stretchability than conventional cross-linked linear polymers with the same degree of polymerization. The superior stretchability is related to both the absence of an entanglement network in ring polymers and the more compact conformations of ring polymers in the undeformed state. 

Polymers are modeled with the common bead-spring model \cite{Kremer1990}, which has been used previously to simulate the static and dynamic properties of non-concatenated ring polymer melts \cite{Halverson2011, Halverson2011a}.  All monomers interact via the truncated and shifted Lennard-Jones (LJ) potential with cut-off $r_c=2.5\sigma$, while chains of $N$ monomers each are connected by the FENE bonding potential. A bond bending potential with a stiffness $1.5\epsilon$ is used to adjust the entanglement length of linear chains to $N_e=28$ beads \cite{Everaers2004}.  The entanglement time $ \tau_e \approx 4 \times 10^3 \tau$ \cite{Ge2014}, which is the relaxation time of an entanglement strand of linear chains.  

Ring polymer melts of length $N=400$, $800$, and $1600$, and linear polymer melts of $N=800$ were well equilibrated at temperature $T=1.0\epsilon/k_B$ to a state with monomer number density $\rho = 0.89 {\sigma}^{-3}$. The number of ring polymers is $M=1600$, $1200$ and $600$ for $N=400$, $800$, and $1600$, respectively.  There are $M=1200$ chains in the linear polymer melt with $N=800$. The melt temperature is maintained by a Nos\'{e}-Hoover thermostat with a characteristic damping time of $1\tau$.

\begin{figure}[tb]
	\includegraphics[width=1.0\columnwidth]{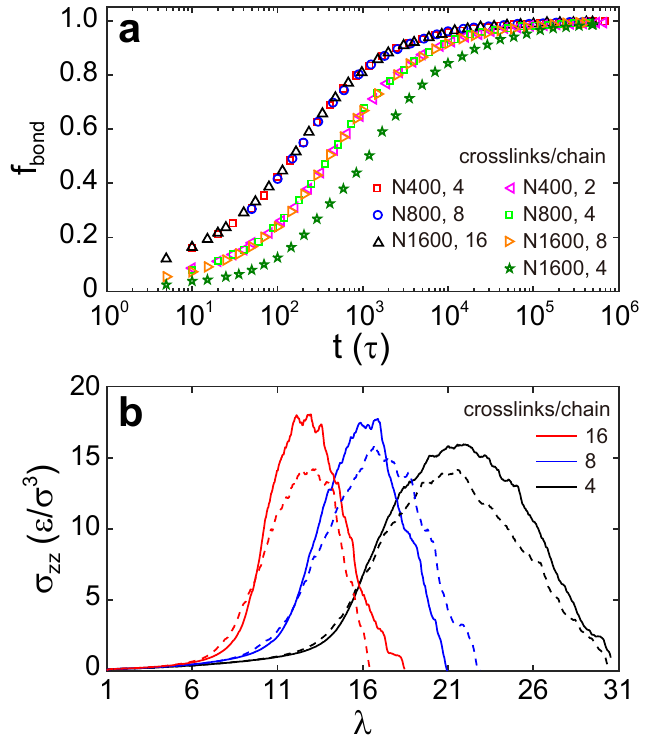}
	\caption{(a) $f_{bond}$ as a function of time $t$ for ring polymers of length $N$ and number of cross-links per chain listed in inset.  (b) Tensile stress $\sigma_{zz}$ as a function of stretch ratio $\lambda$ for randomly cross-linked rings (dashed lines) and kinetically cross-linked rings with regularly spaced cross-linkable monomers (solid lines) for $N=1600$. The red, blue, and black lines are for $16$, $8$, and $4$ cross-links per chain, respectively.}
	\label{fig:fig1}
\end{figure}

To demonstrate the robustness of ring elastomer properties, we use two methods to cross-link the ring polymer melts.  In one method, we randomly pick two monomers that are separated by less than $1.1\sigma$ and link them by a FENE bond. We make sure that any two cross-links are separated by no less than $10$ bonds along the contour of the same ring.  In the other method,  we pre-select monomers that are regularly distributed along the contour of a ring polymer and allow FENE bonds to kinetically form between two such monomers over time.  Each of these cross-linkable monomers can form at most one new bond with another cross-linkable monomer,  when the distance between two cross-linkable monomers is smaller than $1.3\sigma$.  We track $f_{bond}$, the ratio of the number of newly created bonds to the maximum number of bonds that may be created.  It increases with time as shown in Figure 1a.  Notably, the time dependence of $f_{bond}$ does not depend on the length $N$ of ring polymers.  Instead, it depends only on the number of monomers $S$ between two cross-links of a polymer.  Here the self-similar dynamics of entangled ring polymers serve as a practical advantage and facilitate efficient cross-linking even when polymer chains are long.  In contrast, the cross-linking kinetics of entangled linear polymers would slow dramatically with increasing $N$ due to the reptation dynamics. 

We characterize the mechanics of the cross-linked elastomers by deforming them in uniaxial elongation. To enable the mechanical failure of backbone bonds at large strain, we replace the unbreakable FENE bonds with breakable quartic bonds \cite{Stevens2001, Rottler2002, Ge2013, Wang2021}.  The equilibrium bond length of quartic bonds is the same as that of FENE bonds. As for the FENE bonds, the quartic bonds prevent the polymers from crossing and preserve the topological constraints, but quartic bonds break at a tension $240 \epsilon/\sigma$, or $\sim100$ larger than the van der Waals forces.  The uniaxial elongation is performed at a constant strain rate in the $z$-direction, while the stress components $\sigma_{xx}$ and $\sigma_{yy}$ in the $x$- and $y$-directions are kept zero using a Nos\'{e}-Hoover barostat with a characteristic time $10\tau$. During the deformation, the temperature is maintained at $T=1.0\epsilon/k_B$ using a Nos\'{e}-Hoover thermostat with a characteristic time $1\tau$. 

The tensile stress $\sigma_{zz}$ as a function of the stretch ratio $\lambda$, which is the box size $L_z$ along the $z$-direction divided by the value $L_z^0$ before the deformation, is plotted in Figure 1b for different systems. Each stress-strain curve consists of an initial linear elastic regime, a regime of nonlinear stress increase, and a decrease of the stress after reaching the ultimate strength.  At the same density of cross-links, the stress-strain curves of the randomly cross-linked and kinetically cross-linked systems are similar.  In the rest of the paper, we present the results only for the randomly cross-linked systems.

To reveal the effects of polymer topology on the stretchability of cross-linked polymers, we randomly cross-linked entangled linear polymer melts, and compare the behaviors of cross-linked linear and ring polymers. The number of monomers $S$ between two cross-links in the ring and linear polymers with $N=800$ and $8$ cross-links per polymer are compared in Figure S1 of the Supplemental Material (SM). For the ring polymers, the average $\left< S \right>=100$ as expected for $8$ cross-links per polymer of $N =800$.  For the linear polymers, $\left< S \right>=89$ due to the presence of dangling chain ends. The absence of dangling ends is one advantage of cross-linked ring polymers compared with the cross-linked linear polymers in conventional elastomers.

\begin{figure}[b]
	\includegraphics[width=1.0\columnwidth]{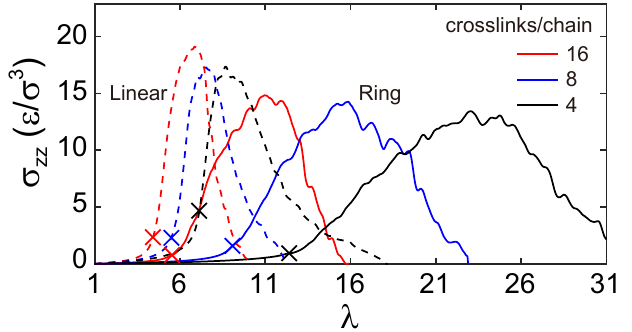}
	\caption{Tensile stress $\sigma_{zz}$ as a function of stretch ratio $\lambda$ for randomly cross-linked ring (solid lines) and linear polymer melts (dashed lines) with $N=800$. The red, blue, and black lines are for $16$, $8$, and $4$ crosslinks per chain, respectively. The cross symbols indicate $\lambda$ at which the bonds begin to break. Deformation rate is $10^{-4}{\tau}^{-1}$.}
	\label{fig:fig2}
\end{figure} 

Figure 2 compares the stress-strain curves of the cross-linked ring and linear polymers with $N=800$ at strain rate $10^{-4}\tau^{-1}$. The points where bonds start to break are indicated by the cross symbols. The subsequent increase in the fraction of broken bonds out of all bonds $f_{broken}$ is shown in Supplemental Figure S2. At the same cross-linking density, the ring polymers are significantly more stretchable than the linear polymers, as reflected in the larger values of $\lambda$ for the end of the initial elastic regime, the emergence of broken bonds, and the stress peak for the ultimate strength.  Supplemental Figure S3 shows that the stretchability of cross-linked ring polymers slightly changes as the deformation rate varies by two orders of magnitude.  The highest rate $10^{-3}{\tau}^{-1}$ is $4$ times the inverse of the entanglement time $\tau_e$,  while the slower rates $10^{-4}{\tau}^{-1}$ and $10^{-5}{\tau}^{-1}$ are both smaller than $\tau_e^{-1}$, allowing the adjustment of entanglements. 

Gels of cross-linked cyclic polymers have been synthesized in experiments and compared with the gels of cross-linked linear polymers in terms of both swelling behavior and mechanical properties \cite{Zhang2011, Trachsel2021}. The higher swelling ratio and the higher strain at the breaking point are consistent with the observation of the higher stretchability of cross-linked ring polymers in our simulations. 

As the cross-linking density decreases from $16$ cross-links per polymer to $4$ cross-links per polymer, the increase of the stretchability is more prominent in the ring polymers compared to that in the linear polymers. This suggests that the stretchability of cross-linked linear polymers is controlled by the trapped entanglements \cite{Pearson1980,Duering1994} and thus only weakly affected by the reduction in the cross-linking density.  By contrast, the stretchability of cross-linked ring polymers is controlled by the strands between cross-links and only depends on the cross-linking density.  

According to the elasticity theory of cross-linked linear polymers \cite{Rubinstein2003},  the network shear modulus $G$ is proportional to the sum of the number densities of cross-links and entanglements, which act as effective cross-links, i.e., $G \sim (1/\left<S\right>+1/N_e)$.  For the cross-linked linear polymers of $N=800$,  following the standard protocol established for the entropic elasticity of Gaussian linear chains,  we extract $G$ from the linear regime of the stress-strain curve as $G = \sigma_{zz} / \left(\lambda^2-1/\lambda\right)$.  Figure 3a shows $G$ as a function of $1/\left< S \right>$.  As expected, $G$ scales linearly with $1/\left<S\right>$. The value $G_e = 0.017\epsilon/\sigma^3$ at vanishing $1/\left<S\right>$ corresponds to the contribution from the entanglement network.  At the highest cross-linking density of $16$ cross-links per chain with $1/\left<S\right> = 1/50 = 0.02$, $G_e$ is about $1/2$ of $G$.  At lower cross-linking densities, $G_e$ is more than $1/2$ of $G$ and dominates over the contribution from the cross-links.  A similar linear dependence of $G$ on the inverse of network strand length was observed in the simulations of end-linked polymer networks by Duering \textit{et al.} \cite{Duering1994}.

\begin{figure}[t]
	\includegraphics[width=1.0\columnwidth]{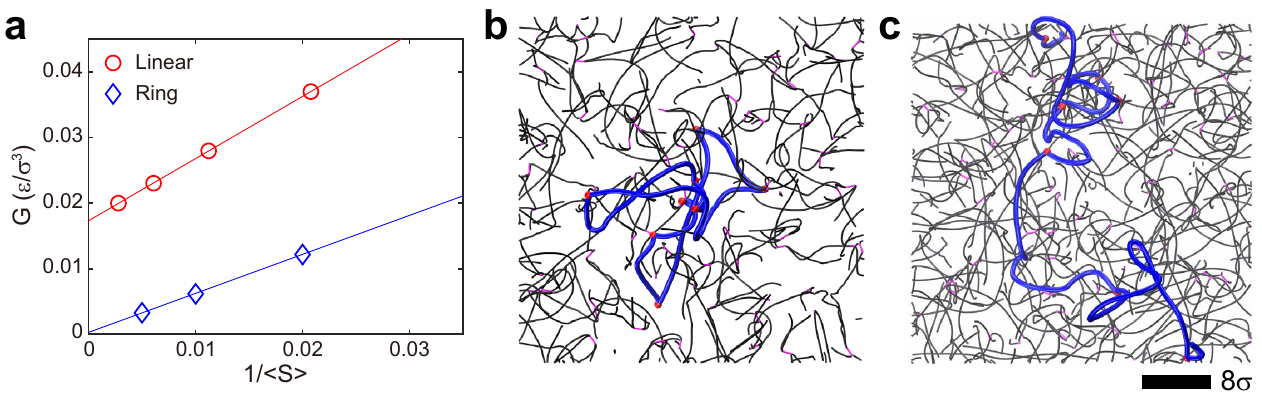}
	\caption{(a) Shear modulus $G$ as a function of $1/\left<S\right>$ for randomly cross-linked  linear and ring polymers of length $N=800$. Snapshots of randomly cross-linked (b) ring and (c) linear polymers of $N=800$ with monomer positions averaged over a period of $4 \times 10^5 \tau$ for a region of size $40 \sigma \times 40 \sigma \times 10 \sigma$. The cross-links ($8$ cross-links per chain) are fixed during the simulations to obtain the averaged monomer positions. One polymer (blue) and the associated cross-links (red spheres) are highlighted, while other polymers (grey) and the associated cross-links (magenta) are shown as a dimmed background.}
	\label{fig:fig3}
\end{figure} 

Despite of the loopy globular conformation,  the end-to-end vector of a section in a ring polymer also follows the Gaussian distribution,  as shown in Supplemental Figure 4. Therefore,  the shear modulus $G$ may be extracted using the same protocol for linear chains.  As shown in Figure 3a,  $G$ for cross-linked rings also scales linearly with $1/\left<S\right>$,  but approaches $0$ at vanishing $1/\left<S\right>$ due to the absence of an entanglement network. At all cross-linking densities, $G$ is solely determined by the cross-links with no entanglements acting as effective cross-links. This result is consistent with the absence of a rubbery plateau in the stress relaxation function for the non-concatenated ring polymer melts studied here \cite{Halverson2011a}. It also demonstrates that the strain rate in the simulations is sufficiently slow for a response near the thermal equilibrium. Supplemental Figure S5 shows that there is a finite modulus at vanishing $1/\left<S\right>$ for a higher strain rate $10^{-3}\tau^{-1}$, at which some entanglements contribute to $G$ as a response out of equilibrium \cite{note1}. 

Generally,  the shear modulus and the stretchability are inversely correlated for cross-linked linear polymers \cite{Sheiko2019}. This inverse correlation is also observed for the cross-linked ring polymers. Along with the super-stretchability, the cross-linked ring polymers exhibit super-softness. As shown in Figure 3a, the modulus of cross-linked ring polymers, which is controlled by the cross-linking density, is significantly lower than that of cross-linked linear polymers and may be ultra-low.

The contrasting states of entanglements in the cross-linked ring and linear polymers are visualized in Figure 3b and Figure 3c. The entanglements between cross-linked polymers are inherited from those in the polymer melt. Time-averaging the positions of monomers with the cross-links fixed in space reveals the underlying entanglements. The time-averaging over a period of $4 \times 10^5 \tau \approx 100\tau_e$ leads to a dense entanglement network in the linear polymers, as the dynamics of entangled linear polymers are confined in tube-like regions \cite{Rubinstein2003}.  By contrast,  after the same period of time-averaging, there is a sparser network in the ring polymers, as the self-similar dynamics of entangled ring polymers progressively reduce the number of entanglements.

On the molecular level,  a polymer chain becomes taut upon stretching.  A visual comparison of the stretching of one polymer in the cross-linked ring and linear polymers is shown in Figure 4 for $N=800$.  At $\lambda=1$, the ring polymers are more compact than the linear polymers. The three values of $\lambda>1$ shown correspond to the end of the linear elastic regime, the emergence of broken bonds, and at the stress peak. 

\begin{figure}[t]
	\includegraphics[width=1.0\columnwidth]{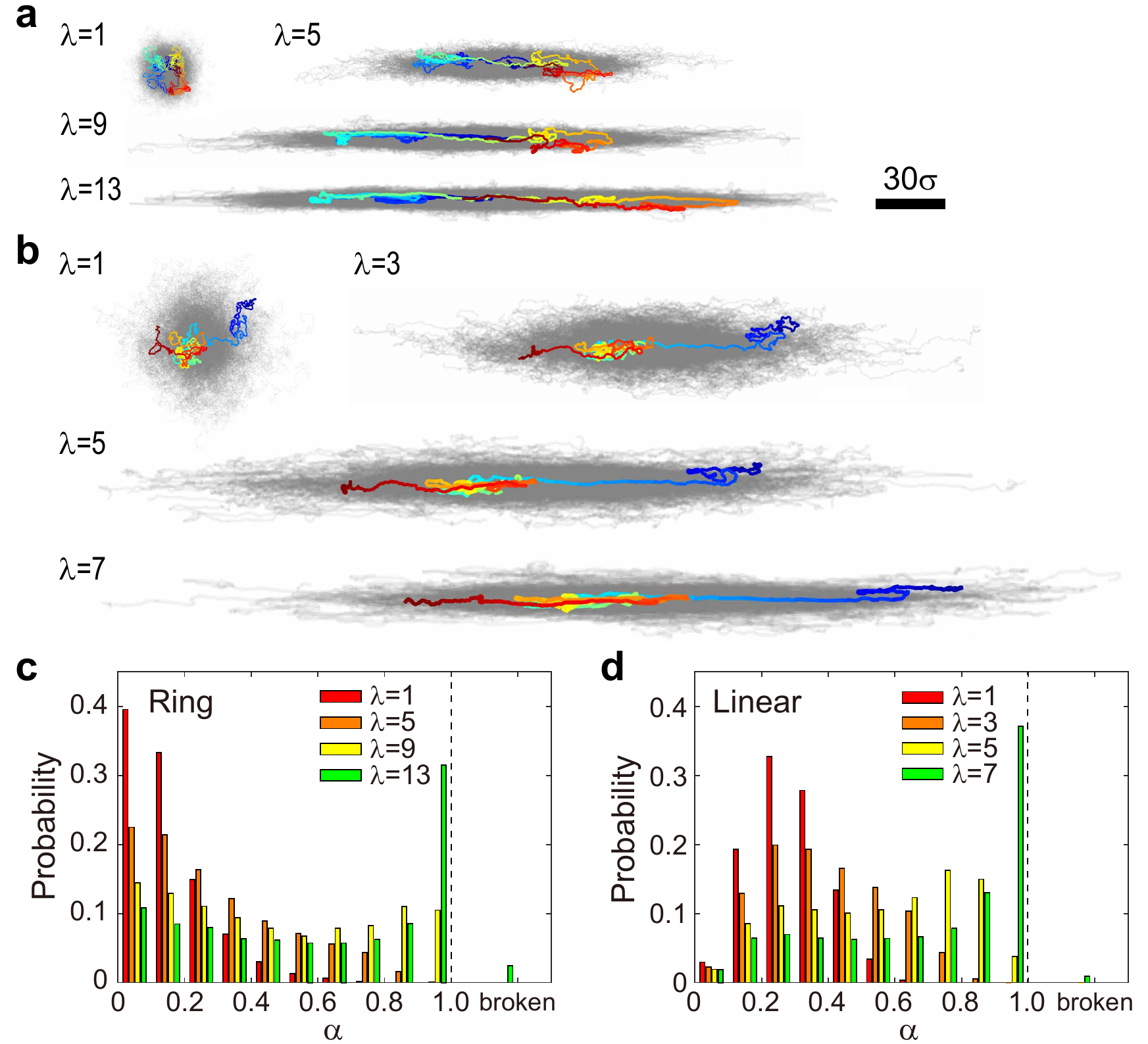}
	\caption{Snapshots of randomly cross-linked (a) ring and (b) linear polymer melts of $N=800$ with $8$ cross-links per chain at indicated $\lambda$. The centers of mass of the polymers are shifted to overlap. One polymer is highlighted with the monomers colored based on their positions along the polymer contour. The radius of gyration of the highlighted polymer is closest to the mean value for all the polymers at $\lambda=1$.  Probability distributions of the degree of tautness $\alpha$ for (c) the strands between cross-links in the ring polymers and (d) the entanglement strands of $N_e = 28$ in the linear polymers.  The values of $\lambda$ in (c) and (d) are identical to those in (a) and (b), respectively.}
	\label{fig:fig4}
\end{figure}

For cross-linked linear polymers,  the stretchability is determined by the maximum extension of an entanglement strand $\lambda_{max}^L = l_e/d_e = N_e^{1/2}/C_{\infty}^{1/2}$,  
%\begin{equation}
%\lambda_{max}^L = l_e/d_e = \frac {N_e l_0} {(C_{\infty}N_e l_0^2)^{1/2}} = \frac{N_e^{1/2}}{C_{\infty}^{1/2}},
%\label{eqn:Eq 1}
%\end{equation}
where $l_e = N_e l_0$ and $d_e = (C_{\infty}N_e l_0^2)^{1/2}$ are the contour length and the root-mean-squared (RMS) end-to-end distance of an entanglement strand as an ideal random-walk chain with average bond length $l_0$ and characteristic ratio $C_{\infty}$.  For $N_e=28$ and $C_{\infty}=2.8$, $\lambda_{max}^L=3.2$.  In the underformed state,  the end-to-end vectors of entanglement strands are randomly orientated with respect to the stretching direction.  As a result, only a small fraction of entanglement strands with their initial orientations close to the stretching direction can reach the maximum extension at $\lambda = \lambda_{max}^L$.  This is demonstrated by the probability distribution of the degree of tautness $\alpha^{L} = d_e / l_e$ of the entanglement strands at $\lambda = 3$ in Figure 4d.  Despite the low fraction of taut entanglement strands,  the non-linear regime of the stress-strain curve initiates around $\lambda = \lambda_{max}^L$ (Figure 2).  With further stretching,  more entanglement strands become taut and more bonds break, as shown in Figure 4d.  Since the average projection of initial entanglement strands to the stretching direction is $d_e /\sqrt{3}$,  $\lambda_p^{L}$ at the stress peak is estimated as $\sqrt{3} \lambda_{max}^L = 5.5$.  For the three cross-linked linear polymers in Figure 2, $\lambda_p^{L} \approx 6.9$, $7.5$, and $8.7$.

For cross-linked ring polymers,  the stretchability is controlled by the maximum extension of a strand between cross-links 
\begin{equation}
\lambda_{max}^R = \frac{l_s}{d_s} = \frac{{(S/N_e)}^{2/3} N_e^{1/2}}{C_{\infty}^{1/2}},
\label{eqn:Eq 1}
\end{equation}
where $l_s = S l_0$ and $d_s = {(S/N_e)}^{1/3} (C_{\infty} N_e l_0^2)^{1/2}$ are the contour length and RMS end-to-end distance of a network strand in the loopy globular conformation.  For $\left<S\right>=100$,  $\lambda_{max}^R = 7.4$.  At $\lambda \approx \lambda_{max}^R$,  only a small fraction of the network strands are taut, as shown by the probability distribution of the degree of tautness $\alpha^{R} = d_s /l_s$ in Figure 4c. However, the non-linear regime begins at $\lambda \approx \lambda_{max}^R$, as shown in Figure 2.  The stretch ratio at the stress peak is estimated as $\lambda_p^{R} = \sqrt{3} \lambda_{max}^{R} = 12.8$. As a comparison,  $\lambda_p^{R} \approx 15.9$ for $8$ cross-links per chain in Figure 2.  At the other two cross-linking densities with $\left<S\right>=50$ and $\left<S\right>=200$,  $\lambda_{max}^R=4.7$ and $11.7$,  both corresponding to $\lambda$ at the beginning of the non-linear regime in Figure 2.  The estimated $\lambda_p^{R} = \sqrt{3}\lambda_{max}^{R}=8.1$ and $20.3$ for the two systems,  and the results in Figure 2 are $\lambda_p^{R} \approx 11.0$ and $23.0$. 

The increase in the stretchability of cross-linked ring polymers with respect to that of the linear counterpart is reflected in the ratio
\begin{equation}
\frac{\lambda_{max}^R}{\lambda_{max}^L} = \left(\frac{S}{N_e}\right)^{2/3},
\label{eqn:Eq 2}
\end{equation}
which is independent of the local stiffness as quantified by $C_{\infty}$.  If the network strand of cross-linked ring polymers were an ideal random-walk (RW) chain, the maximum extension $\lambda_{max}^{R, RW} = l_s/d_s =  S^{1/2}/C_{\infty}^{1/2}$ is larger than $\lambda_{max}^L$ only by $(S / N_e)^{1/2}$. The factor 
$\lambda_{max}^R / \lambda_{max}^{R,RW} = (S / N_e)^{1/6}$
quantifies the additional contribution to the higher stretchability of ring polymers from the more compact conformations before stretching. %The low degree of tautness for the strands between cross-links at $\lambda = 1$ in Figure 4c reflects the compactness of ring polymers.  

Here we have shown that cross-linked ring polymers are a viable and robust route to generate super-stretchable elastomers.  
Unlike linear chains, non-concatenated rings do not form a network of trapped entanglements even when $N$ is large, allowing ring elastomers to overcome the entanglement barrier that limits the stretchability of other elastomer architectures.  
Ring cross-linking kinetics are strikingly $N$-independent, even in concentrated melts.
This enables the efficient creation of elastomers from much higher molecular weight rings than could be achieved for linear chains.
We have shown that these superior properties are rooted in the loopy globular conformations and self-similar dynamics of the non-concatenated rings.
Given recent advances in reversible polymerization of non-concantenated ring polymers, we believe our results provide a timely and compelling demonstration of how ring polymers could be used to create practical and sustainable materials \cite{Romio2020}.
We suspect these materials could be made even more stretchable by incorporating reversible and exchangeable cross-links within the network \cite{Fu2018, Guo2020, Sun2012}.

\begin{acknowledgments}	
T. Ge acknowledges start-up funds from the University of South Carolina. This work was supported in part by the National Science Foundation EPSCoR Program under NSF Award No. OIA-1655740. Any Opinions, findings and conclusions or recommendations expressed in this material are those of the authors and do not necessarily reflect those of the National Science Foundation. Computational resources were provided by University of South Carolina flagship computing cluster Hyperion. This research used resources at the National Energy Research Scientific Computing Center (NERSC), a U.S. Department of Energy Office of Science User Facility operated under Contract No. DE-AC02-05CH11231. These resources were obtained through the Advanced Scientific Computing Research (ASCR) Leadership Computing Challenge (ALCC). This work was performed, in part, at the Center for Integrated Nanotechnologies, an Office of Science User Facility operated for the U.S. Department of Energy (DOE) Office of Science. Sandia National Laboratories is a multimission laboratory managed and operated by National Technology $\&$ Engineering Solutions of Sandia, LLC, a wholly owned subsidiary of Honeywell International, Inc., for the U.S. DOEs National Nuclear Security Administration under Contract No. DE-NA-0003525. The views expressed in the Letter do not necessarily represent the views of the U.S. DOE or the United States Government.
\end{acknowledgments}

% \bibliography{ring}

\begin{thebibliography}{39}%
\makeatletter
\providecommand \@ifxundefined [1]{%
 \@ifx{#1\undefined}
}%
\providecommand \@ifnum [1]{%
 \ifnum #1\expandafter \@firstoftwo
 \else \expandafter \@secondoftwo
 \fi
}%
\providecommand \@ifx [1]{%
 \ifx #1\expandafter \@firstoftwo
 \else \expandafter \@secondoftwo
 \fi
}%
\providecommand \natexlab [1]{#1}%
\providecommand \enquote  [1]{``#1''}%
\providecommand \bibnamefont  [1]{#1}%
\providecommand \bibfnamefont [1]{#1}%
\providecommand \citenamefont [1]{#1}%
\providecommand \href@noop [0]{\@secondoftwo}%
\providecommand \href [0]{\begingroup \@sanitize@url \@href}%
\providecommand \@href[1]{\@@startlink{#1}\@@href}%
\providecommand \@@href[1]{\endgroup#1\@@endlink}%
\providecommand \@sanitize@url [0]{\catcode `\\12\catcode `\$12\catcode
  `\&12\catcode `\#12\catcode `\^12\catcode `\_12\catcode `\%12\relax}%
\providecommand \@@startlink[1]{}%
\providecommand \@@endlink[0]{}%
\providecommand \url  [0]{\begingroup\@sanitize@url \@url }%
\providecommand \@url [1]{\endgroup\@href {#1}{\urlprefix }}%
\providecommand \urlprefix  [0]{URL }%
\providecommand \Eprint [0]{\href }%
\providecommand \doibase [0]{http://dx.doi.org/}%
\providecommand \selectlanguage [0]{\@gobble}%
\providecommand \bibinfo  [0]{\@secondoftwo}%
\providecommand \bibfield  [0]{\@secondoftwo}%
\providecommand \translation [1]{[#1]}%
\providecommand \BibitemOpen [0]{}%
\providecommand \bibitemStop [0]{}%
\providecommand \bibitemNoStop [0]{.\EOS\space}%
\providecommand \EOS [0]{\spacefactor3000\relax}%
\providecommand \BibitemShut  [1]{\csname bibitem#1\endcsname}%
\let\auto@bib@innerbib\@empty
%</preamble>
\bibitem [{\citenamefont {Gul}\ \emph {et~al.}(2018)\citenamefont {Gul},
  \citenamefont {Sajid}, \citenamefont {Rehman}, \citenamefont {Siddiqui},
  \citenamefont {Shah}, \citenamefont {Kim}, \citenamefont {Lee},\ and\
  \citenamefont {Choi}}]{Gul2018}%
  \BibitemOpen
  \bibfield  {author} {\bibinfo {author} {\bibfnamefont {J.~Z.}\ \bibnamefont
  {Gul}}, \bibinfo {author} {\bibfnamefont {M.}~\bibnamefont {Sajid}}, \bibinfo
  {author} {\bibfnamefont {M.~M.}\ \bibnamefont {Rehman}}, \bibinfo {author}
  {\bibfnamefont {G.~U.}\ \bibnamefont {Siddiqui}}, \bibinfo {author}
  {\bibfnamefont {I.}~\bibnamefont {Shah}}, \bibinfo {author} {\bibfnamefont
  {K.-H.}\ \bibnamefont {Kim}}, \bibinfo {author} {\bibfnamefont {J.-W.}\
  \bibnamefont {Lee}}, \ and\ \bibinfo {author} {\bibfnamefont {K.~H.}\
  \bibnamefont {Choi}},\ }\href@noop {} {\bibfield  {journal} {\bibinfo
  {journal} {Sci. Technol. Adv. Mater.}\ }\textbf {\bibinfo {volume} {19}},\
  \bibinfo {pages} {243} (\bibinfo {year} {2018})}\BibitemShut {NoStop}%
\bibitem [{\citenamefont {Amjadi}\ \emph {et~al.}(2016)\citenamefont {Amjadi},
  \citenamefont {Kyung}, \citenamefont {Park},\ and\ \citenamefont
  {Sitti}}]{Amjadi2016}%
  \BibitemOpen
  \bibfield  {author} {\bibinfo {author} {\bibfnamefont {M.}~\bibnamefont
  {Amjadi}}, \bibinfo {author} {\bibfnamefont {K.}~\bibnamefont {Kyung}},
  \bibinfo {author} {\bibfnamefont {I.}~\bibnamefont {Park}}, \ and\ \bibinfo
  {author} {\bibfnamefont {M.}~\bibnamefont {Sitti}},\ }\href@noop {}
  {\bibfield  {journal} {\bibinfo  {journal} {Adv. Funct. Mater.}\ }\textbf
  {\bibinfo {volume} {26}},\ \bibinfo {pages} {1678} (\bibinfo {year}
  {2016})}\BibitemShut {NoStop}%
\bibitem [{\citenamefont {Wu}(2019)}]{Wu2019}%
  \BibitemOpen
  \bibfield  {author} {\bibinfo {author} {\bibfnamefont {W.}~\bibnamefont
  {Wu}},\ }\href@noop {} {\bibfield  {journal} {\bibinfo  {journal} {Sci.
  Technol. Adv. Mater.}\ }\textbf {\bibinfo {volume} {20}},\ \bibinfo {pages}
  {187} (\bibinfo {year} {2019})}\BibitemShut {NoStop}%
\bibitem [{\citenamefont {Yeo}\ and\ \citenamefont {Lim}(2016)}]{Yeo2016}%
  \BibitemOpen
  \bibfield  {author} {\bibinfo {author} {\bibfnamefont {J.~C.}\ \bibnamefont
  {Yeo}}\ and\ \bibinfo {author} {\bibfnamefont {C.~T.}\ \bibnamefont {Lim}},\
  }\href@noop {} {\bibfield  {journal} {\bibinfo  {journal} {Microsyst.
  Nanoeng.}\ }\textbf {\bibinfo {volume} {2}},\ \bibinfo {pages} {1} (\bibinfo
  {year} {2016})}\BibitemShut {NoStop}%
\bibitem [{\citenamefont {Pearson}\ and\ \citenamefont
  {Graessley}(1980)}]{Pearson1980}%
  \BibitemOpen
  \bibfield  {author} {\bibinfo {author} {\bibfnamefont {D.~S.}\ \bibnamefont
  {Pearson}}\ and\ \bibinfo {author} {\bibfnamefont {W.~W.}\ \bibnamefont
  {Graessley}},\ }\href@noop {} {\bibfield  {journal} {\bibinfo  {journal}
  {Macromolecules}\ }\textbf {\bibinfo {volume} {13}},\ \bibinfo {pages} {1001}
  (\bibinfo {year} {1980})}\BibitemShut {NoStop}%
\bibitem [{\citenamefont {Gu}\ \emph {et~al.}(2020)\citenamefont {Gu},
  \citenamefont {Zhao},\ and\ \citenamefont {Johnson}}]{Gu2020}%
  \BibitemOpen
  \bibfield  {author} {\bibinfo {author} {\bibfnamefont {Y.}~\bibnamefont
  {Gu}}, \bibinfo {author} {\bibfnamefont {J.}~\bibnamefont {Zhao}}, \ and\
  \bibinfo {author} {\bibfnamefont {J.~A.}\ \bibnamefont {Johnson}},\ }\href
  {\doibase 10.1002/anie.201902900} {\bibfield  {journal} {\bibinfo  {journal}
  {Angew. Chem. Int. Ed.}\ }\textbf {\bibinfo {volume} {59}},\ \bibinfo {pages}
  {5022} (\bibinfo {year} {2020})}\BibitemShut {NoStop}%
\bibitem [{\citenamefont {Cai}\ \emph {et~al.}(2015)\citenamefont {Cai},
  \citenamefont {Kodger}, \citenamefont {Guerra}, \citenamefont {Pegoraro},
  \citenamefont {Rubinstein},\ and\ \citenamefont {Weitz}}]{Cai2015}%
  \BibitemOpen
  \bibfield  {author} {\bibinfo {author} {\bibfnamefont {L.}~\bibnamefont
  {Cai}}, \bibinfo {author} {\bibfnamefont {T.~E.}\ \bibnamefont {Kodger}},
  \bibinfo {author} {\bibfnamefont {R.~E.}\ \bibnamefont {Guerra}}, \bibinfo
  {author} {\bibfnamefont {A.~F.}\ \bibnamefont {Pegoraro}}, \bibinfo {author}
  {\bibfnamefont {M.}~\bibnamefont {Rubinstein}}, \ and\ \bibinfo {author}
  {\bibfnamefont {D.~A.}\ \bibnamefont {Weitz}},\ }\href@noop {} {\bibfield
  {journal} {\bibinfo  {journal} {Adv. Mater.}\ }\textbf {\bibinfo {volume}
  {27}},\ \bibinfo {pages} {5132} (\bibinfo {year} {2015})}\BibitemShut
  {NoStop}%
\bibitem [{\citenamefont {Cao}\ \emph {et~al.}(2015)\citenamefont {Cao},
  \citenamefont {Carrillo}, \citenamefont {Sheiko},\ and\ \citenamefont
  {Dobrynin}}]{Cao2015}%
  \BibitemOpen
  \bibfield  {author} {\bibinfo {author} {\bibfnamefont {Z.}~\bibnamefont
  {Cao}}, \bibinfo {author} {\bibfnamefont {J.-M.~Y.}\ \bibnamefont
  {Carrillo}}, \bibinfo {author} {\bibfnamefont {S.~S.}\ \bibnamefont
  {Sheiko}}, \ and\ \bibinfo {author} {\bibfnamefont {A.~V.}\ \bibnamefont
  {Dobrynin}},\ }\href@noop {} {\bibfield  {journal} {\bibinfo  {journal}
  {Macromolecules}\ }\textbf {\bibinfo {volume} {48}},\ \bibinfo {pages} {5006}
  (\bibinfo {year} {2015})}\BibitemShut {NoStop}%
\bibitem [{\citenamefont {Daniel}\ \emph {et~al.}(2016)\citenamefont {Daniel},
  \citenamefont {Burdy{\'{n}}ska}, \citenamefont {Vatankhah-varnoosfaderani},
  \citenamefont {Matyjaszewski}, \citenamefont {Paturej}, \citenamefont
  {Rubinstein}, \citenamefont {Dobrynin},\ and\ \citenamefont
  {Sheiko}}]{Daniel2016}%
  \BibitemOpen
  \bibfield  {author} {\bibinfo {author} {\bibfnamefont {W.~F.~M.}\
  \bibnamefont {Daniel}}, \bibinfo {author} {\bibfnamefont {J.}~\bibnamefont
  {Burdy{\'{n}}ska}}, \bibinfo {author} {\bibfnamefont {M.}~\bibnamefont
  {Vatankhah-varnoosfaderani}}, \bibinfo {author} {\bibfnamefont
  {K.}~\bibnamefont {Matyjaszewski}}, \bibinfo {author} {\bibfnamefont
  {J.}~\bibnamefont {Paturej}}, \bibinfo {author} {\bibfnamefont
  {M.}~\bibnamefont {Rubinstein}}, \bibinfo {author} {\bibfnamefont {A.~V.}\
  \bibnamefont {Dobrynin}}, \ and\ \bibinfo {author} {\bibfnamefont {S.~S.}\
  \bibnamefont {Sheiko}},\ }\href {\doibase 10.1038/nmat4508} {\bibfield
  {journal} {\bibinfo  {journal} {Nat. Mater.}\ }\textbf {\bibinfo {volume}
  {15}},\ \bibinfo {pages} {183} (\bibinfo {year} {2016})}\BibitemShut
  {NoStop}%
\bibitem [{\citenamefont {Sheiko}\ and\ \citenamefont
  {Dobrynin}(2019)}]{Sheiko2019}%
  \BibitemOpen
  \bibfield  {author} {\bibinfo {author} {\bibfnamefont {S.~S.}\ \bibnamefont
  {Sheiko}}\ and\ \bibinfo {author} {\bibfnamefont {A.~V.}\ \bibnamefont
  {Dobrynin}},\ }\href@noop {} {\bibfield  {journal} {\bibinfo  {journal}
  {Macromolecules}\ }\textbf {\bibinfo {volume} {52}},\ \bibinfo {pages} {7531}
  (\bibinfo {year} {2019})}\BibitemShut {NoStop}%
\bibitem [{\citenamefont {Obukhov}\ \emph {et~al.}(2014)\citenamefont
  {Obukhov}, \citenamefont {Johner}, \citenamefont {Baschnagel}, \citenamefont
  {Meyer},\ and\ \citenamefont {Wittmer}}]{Obukhov2014}%
  \BibitemOpen
  \bibfield  {author} {\bibinfo {author} {\bibfnamefont {S.}~\bibnamefont
  {Obukhov}}, \bibinfo {author} {\bibfnamefont {A.}~\bibnamefont {Johner}},
  \bibinfo {author} {\bibfnamefont {J.}~\bibnamefont {Baschnagel}}, \bibinfo
  {author} {\bibfnamefont {H.}~\bibnamefont {Meyer}}, \ and\ \bibinfo {author}
  {\bibfnamefont {J.}~\bibnamefont {Wittmer}},\ }\href {\doibase Artn 48005
  10.1209/0295-5075/105/48005} {\bibfield  {journal} {\bibinfo  {journal}
  {Europhys. Lett.}\ }\textbf {\bibinfo {volume} {105}},\ \bibinfo {pages}
  {48005} (\bibinfo {year} {2014})}\BibitemShut {NoStop}%
\bibitem [{\citenamefont {Goo{\ss}en}\ \emph {et~al.}(2014)\citenamefont
  {Goo{\ss}en}, \citenamefont {Br{\'a}s}, \citenamefont {Krutyeva},
  \citenamefont {Sharp}, \citenamefont {Falus}, \citenamefont {Feoktystov},
  \citenamefont {Gasser}, \citenamefont {Pyckhout-Hintzen}, \citenamefont
  {Wischnewski},\ and\ \citenamefont {Richter}}]{Gooben2014}%
  \BibitemOpen
  \bibfield  {author} {\bibinfo {author} {\bibfnamefont {S.}~\bibnamefont
  {Goo{\ss}en}}, \bibinfo {author} {\bibfnamefont {A.~R.}\ \bibnamefont
  {Br{\'a}s}}, \bibinfo {author} {\bibfnamefont {M.}~\bibnamefont {Krutyeva}},
  \bibinfo {author} {\bibfnamefont {M.}~\bibnamefont {Sharp}}, \bibinfo
  {author} {\bibfnamefont {P.}~\bibnamefont {Falus}}, \bibinfo {author}
  {\bibfnamefont {A.}~\bibnamefont {Feoktystov}}, \bibinfo {author}
  {\bibfnamefont {U.}~\bibnamefont {Gasser}}, \bibinfo {author} {\bibfnamefont
  {W.}~\bibnamefont {Pyckhout-Hintzen}}, \bibinfo {author} {\bibfnamefont
  {A.}~\bibnamefont {Wischnewski}}, \ and\ \bibinfo {author} {\bibfnamefont
  {D.}~\bibnamefont {Richter}},\ }\href {\doibase
  10.1103/PhysRevLett.113.168302} {\bibfield  {journal} {\bibinfo  {journal}
  {Phys. Rev. Lett.}\ }\textbf {\bibinfo {volume} {113}},\ \bibinfo {pages}
  {168302} (\bibinfo {year} {2014})}\BibitemShut {NoStop}%
\bibitem [{\citenamefont {Rosa}\ and\ \citenamefont
  {Everaers}(2014)}]{Rosa2014}%
  \BibitemOpen
  \bibfield  {author} {\bibinfo {author} {\bibfnamefont {A.}~\bibnamefont
  {Rosa}}\ and\ \bibinfo {author} {\bibfnamefont {R.}~\bibnamefont
  {Everaers}},\ }\href {\doibase 10.1103/PhysRevLett.112.118302} {\bibfield
  {journal} {\bibinfo  {journal} {Phys. Rev. Lett.}\ }\textbf {\bibinfo
  {volume} {112}},\ \bibinfo {pages} {118302} (\bibinfo {year}
  {2014})}\BibitemShut {NoStop}%
\bibitem [{\citenamefont {Ge}\ \emph {et~al.}(2016)\citenamefont {Ge},
  \citenamefont {Panyukov},\ and\ \citenamefont {Rubinstein}}]{Ge2016}%
  \BibitemOpen
  \bibfield  {author} {\bibinfo {author} {\bibfnamefont {T.}~\bibnamefont
  {Ge}}, \bibinfo {author} {\bibfnamefont {S.}~\bibnamefont {Panyukov}}, \ and\
  \bibinfo {author} {\bibfnamefont {M.}~\bibnamefont {Rubinstein}},\ }\href
  {\doibase 10.1021/acs.macromol.5b02319} {\bibfield  {journal} {\bibinfo
  {journal} {Macromolecules}\ }\textbf {\bibinfo {volume} {49}},\ \bibinfo
  {pages} {708} (\bibinfo {year} {2016})}\BibitemShut {NoStop}%
\bibitem [{\citenamefont {Kruteva}\ \emph
  {et~al.}(2020{\natexlab{a}})\citenamefont {Kruteva}, \citenamefont
  {Allgaier}, \citenamefont {Monkenbusch}, \citenamefont {Porcar},\ and\
  \citenamefont {Richter}}]{Kruteva2020a}%
  \BibitemOpen
  \bibfield  {author} {\bibinfo {author} {\bibfnamefont {M.}~\bibnamefont
  {Kruteva}}, \bibinfo {author} {\bibfnamefont {J.}~\bibnamefont {Allgaier}},
  \bibinfo {author} {\bibfnamefont {M.}~\bibnamefont {Monkenbusch}}, \bibinfo
  {author} {\bibfnamefont {L.}~\bibnamefont {Porcar}}, \ and\ \bibinfo {author}
  {\bibfnamefont {D.}~\bibnamefont {Richter}},\ }\href {\doibase
  10.1021/acsmacrolett.0c00190} {\bibfield  {journal} {\bibinfo  {journal} {ACS
  Macro Lett.}\ }\textbf {\bibinfo {volume} {9}},\ \bibinfo {pages} {507}
  (\bibinfo {year} {2020}{\natexlab{a}})}\BibitemShut {NoStop}%
\bibitem [{\citenamefont {Kruteva}\ \emph
  {et~al.}(2020{\natexlab{b}})\citenamefont {Kruteva}, \citenamefont
  {Monkenbusch}, \citenamefont {Allgaier}, \citenamefont {Holderer},
  \citenamefont {Pasini}, \citenamefont {Hoffmann},\ and\ \citenamefont
  {Richter}}]{Kruteva2020}%
  \BibitemOpen
  \bibfield  {author} {\bibinfo {author} {\bibfnamefont {M.}~\bibnamefont
  {Kruteva}}, \bibinfo {author} {\bibfnamefont {M.}~\bibnamefont
  {Monkenbusch}}, \bibinfo {author} {\bibfnamefont {J.}~\bibnamefont
  {Allgaier}}, \bibinfo {author} {\bibfnamefont {O.}~\bibnamefont {Holderer}},
  \bibinfo {author} {\bibfnamefont {S.}~\bibnamefont {Pasini}}, \bibinfo
  {author} {\bibfnamefont {I.}~\bibnamefont {Hoffmann}}, \ and\ \bibinfo
  {author} {\bibfnamefont {D.}~\bibnamefont {Richter}},\ }\href {\doibase
  10.1103/PhysRevLett.125.238004} {\bibfield  {journal} {\bibinfo  {journal}
  {Phys. Rev. Lett.}\ }\textbf {\bibinfo {volume} {125}},\ \bibinfo {pages}
  {238004} (\bibinfo {year} {2020}{\natexlab{b}})}\BibitemShut {NoStop}%
\bibitem [{\citenamefont {Kapnistos}\ \emph {et~al.}(2008)\citenamefont
  {Kapnistos}, \citenamefont {Lang}, \citenamefont {Vlassopoulos},
  \citenamefont {Pyckhout-Hintzen}, \citenamefont {Richter}, \citenamefont
  {Cho}, \citenamefont {Chang},\ and\ \citenamefont
  {Rubinstein}}]{Kapnistos2008}%
  \BibitemOpen
  \bibfield  {author} {\bibinfo {author} {\bibfnamefont {M.}~\bibnamefont
  {Kapnistos}}, \bibinfo {author} {\bibfnamefont {M.}~\bibnamefont {Lang}},
  \bibinfo {author} {\bibfnamefont {D.}~\bibnamefont {Vlassopoulos}}, \bibinfo
  {author} {\bibfnamefont {W.}~\bibnamefont {Pyckhout-Hintzen}}, \bibinfo
  {author} {\bibfnamefont {D.}~\bibnamefont {Richter}}, \bibinfo {author}
  {\bibfnamefont {D.}~\bibnamefont {Cho}}, \bibinfo {author} {\bibfnamefont
  {T.}~\bibnamefont {Chang}}, \ and\ \bibinfo {author} {\bibfnamefont
  {M.}~\bibnamefont {Rubinstein}},\ }\href {\doibase 10.1038/nmat2292}
  {\bibfield  {journal} {\bibinfo  {journal} {Nat. Mater.}\ }\textbf {\bibinfo
  {volume} {7}},\ \bibinfo {pages} {997} (\bibinfo {year} {2008})}\BibitemShut
  {NoStop}%
\bibitem [{\citenamefont {Pasquino}\ \emph {et~al.}(2013)\citenamefont
  {Pasquino}, \citenamefont {Vasilakopoulos}, \citenamefont {Jeong},
  \citenamefont {Lee}, \citenamefont {Rogers}, \citenamefont {Sakellariou},
  \citenamefont {Allgaier}, \citenamefont {Takano}, \citenamefont {Br{\'{a}}s},
  \citenamefont {Chang}, \citenamefont {Goo{\ss}en}, \citenamefont
  {Pyckhout-Hintzen}, \citenamefont {Wischnewski}, \citenamefont
  {Hadjichristidis}, \citenamefont {Richter}, \citenamefont {Rubinstein},\ and\
  \citenamefont {Vlassopoulos}}]{Pasquino2013}%
  \BibitemOpen
  \bibfield  {author} {\bibinfo {author} {\bibfnamefont {R.}~\bibnamefont
  {Pasquino}}, \bibinfo {author} {\bibfnamefont {T.~C.}\ \bibnamefont
  {Vasilakopoulos}}, \bibinfo {author} {\bibfnamefont {Y.~C.}\ \bibnamefont
  {Jeong}}, \bibinfo {author} {\bibfnamefont {H.}~\bibnamefont {Lee}}, \bibinfo
  {author} {\bibfnamefont {S.}~\bibnamefont {Rogers}}, \bibinfo {author}
  {\bibfnamefont {G.}~\bibnamefont {Sakellariou}}, \bibinfo {author}
  {\bibfnamefont {J.}~\bibnamefont {Allgaier}}, \bibinfo {author}
  {\bibfnamefont {A.}~\bibnamefont {Takano}}, \bibinfo {author} {\bibfnamefont
  {A.~R.}\ \bibnamefont {Br{\'{a}}s}}, \bibinfo {author} {\bibfnamefont
  {T.}~\bibnamefont {Chang}}, \bibinfo {author} {\bibfnamefont
  {S.}~\bibnamefont {Goo{\ss}en}}, \bibinfo {author} {\bibfnamefont
  {W.}~\bibnamefont {Pyckhout-Hintzen}}, \bibinfo {author} {\bibfnamefont
  {A.}~\bibnamefont {Wischnewski}}, \bibinfo {author} {\bibfnamefont
  {N.}~\bibnamefont {Hadjichristidis}}, \bibinfo {author} {\bibfnamefont
  {D.}~\bibnamefont {Richter}}, \bibinfo {author} {\bibfnamefont
  {M.}~\bibnamefont {Rubinstein}}, \ and\ \bibinfo {author} {\bibfnamefont
  {D.}~\bibnamefont {Vlassopoulos}},\ }\href {\doibase 10.1021/mz400344e}
  {\bibfield  {journal} {\bibinfo  {journal} {ACS Macro Lett.}\ }\textbf
  {\bibinfo {volume} {2}},\ \bibinfo {pages} {874} (\bibinfo {year}
  {2013})}\BibitemShut {NoStop}%
\bibitem [{\citenamefont {Huang}\ \emph {et~al.}(2019)\citenamefont {Huang},
  \citenamefont {Ahn}, \citenamefont {Parisi}, \citenamefont {Chang},
  \citenamefont {Hassager}, \citenamefont {Panyukov}, \citenamefont
  {Rubinstein},\ and\ \citenamefont {Vlassopoulos}}]{Huang2019}%
  \BibitemOpen
  \bibfield  {author} {\bibinfo {author} {\bibfnamefont {Q.}~\bibnamefont
  {Huang}}, \bibinfo {author} {\bibfnamefont {J.}~\bibnamefont {Ahn}}, \bibinfo
  {author} {\bibfnamefont {D.}~\bibnamefont {Parisi}}, \bibinfo {author}
  {\bibfnamefont {T.}~\bibnamefont {Chang}}, \bibinfo {author} {\bibfnamefont
  {O.}~\bibnamefont {Hassager}}, \bibinfo {author} {\bibfnamefont
  {S.}~\bibnamefont {Panyukov}}, \bibinfo {author} {\bibfnamefont
  {M.}~\bibnamefont {Rubinstein}}, \ and\ \bibinfo {author} {\bibfnamefont
  {D.}~\bibnamefont {Vlassopoulos}},\ }\href {\doibase
  10.1103/PhysRevLett.122.208001} {\bibfield  {journal} {\bibinfo  {journal}
  {Phys. Rev. Lett.}\ }\textbf {\bibinfo {volume} {122}},\ \bibinfo {pages}
  {208001} (\bibinfo {year} {2019})}\BibitemShut {NoStop}%
\bibitem [{\citenamefont {O'Connor}\ \emph {et~al.}(2020)\citenamefont
  {O'Connor}, \citenamefont {Ge}, \citenamefont {Rubinstein},\ and\
  \citenamefont {Grest}}]{Connor2020}%
  \BibitemOpen
  \bibfield  {author} {\bibinfo {author} {\bibfnamefont {T.~C.}\ \bibnamefont
  {O'Connor}}, \bibinfo {author} {\bibfnamefont {T.}~\bibnamefont {Ge}},
  \bibinfo {author} {\bibfnamefont {M.}~\bibnamefont {Rubinstein}}, \ and\
  \bibinfo {author} {\bibfnamefont {G.~S.}\ \bibnamefont {Grest}},\ }\href
  {\doibase 10.1103/PhysRevLett.124.027801} {\bibfield  {journal} {\bibinfo
  {journal} {Phys. Rev. Lett.}\ }\textbf {\bibinfo {volume} {124}},\ \bibinfo
  {pages} {027801} (\bibinfo {year} {2020})}\BibitemShut {NoStop}%
\bibitem [{\citenamefont {Parisi}\ \emph {et~al.}(2021)\citenamefont {Parisi},
  \citenamefont {Costanzo}, \citenamefont {Jeong}, \citenamefont {Ahn},
  \citenamefont {Chang}, \citenamefont {Vlassopoulos}, \citenamefont
  {Halverson}, \citenamefont {Kremer}, \citenamefont {Ge},\ and\ \citenamefont
  {Rubinstein}}]{Parisi2021}%
  \BibitemOpen
  \bibfield  {author} {\bibinfo {author} {\bibfnamefont {D.}~\bibnamefont
  {Parisi}}, \bibinfo {author} {\bibfnamefont {S.}~\bibnamefont {Costanzo}},
  \bibinfo {author} {\bibfnamefont {Y.}~\bibnamefont {Jeong}}, \bibinfo
  {author} {\bibfnamefont {J.}~\bibnamefont {Ahn}}, \bibinfo {author}
  {\bibfnamefont {T.}~\bibnamefont {Chang}}, \bibinfo {author} {\bibfnamefont
  {D.}~\bibnamefont {Vlassopoulos}}, \bibinfo {author} {\bibfnamefont {J.~D.}\
  \bibnamefont {Halverson}}, \bibinfo {author} {\bibfnamefont {K.}~\bibnamefont
  {Kremer}}, \bibinfo {author} {\bibfnamefont {T.}~\bibnamefont {Ge}}, \ and\
  \bibinfo {author} {\bibfnamefont {M.}~\bibnamefont {Rubinstein}},\ }\href
  {\doibase 10.1021/acs.macromol.0c02839} {\bibfield  {journal} {\bibinfo
  {journal} {Macromolecules}\ }\textbf {\bibinfo {volume} {54}},\ \bibinfo
  {pages} {2811} (\bibinfo {year} {2021})}\BibitemShut {NoStop}%
\bibitem [{\citenamefont {Kremer}\ and\ \citenamefont
  {Grest}(1990)}]{Kremer1990}%
  \BibitemOpen
  \bibfield  {author} {\bibinfo {author} {\bibfnamefont {K.}~\bibnamefont
  {Kremer}}\ and\ \bibinfo {author} {\bibfnamefont {G.~S.}\ \bibnamefont
  {Grest}},\ }\href@noop {} {\bibfield  {journal} {\bibinfo  {journal} {J.
  Chem. Phys.}\ }\textbf {\bibinfo {volume} {92}},\ \bibinfo {pages} {5057}
  (\bibinfo {year} {1990})}\BibitemShut {NoStop}%
\bibitem [{\citenamefont {Halverson}\ \emph
  {et~al.}(2011{\natexlab{a}})\citenamefont {Halverson}, \citenamefont {Lee},
  \citenamefont {Grest}, \citenamefont {Grosberg},\ and\ \citenamefont
  {Kremer}}]{Halverson2011}%
  \BibitemOpen
  \bibfield  {author} {\bibinfo {author} {\bibfnamefont {J.~D.}\ \bibnamefont
  {Halverson}}, \bibinfo {author} {\bibfnamefont {W.~B.}\ \bibnamefont {Lee}},
  \bibinfo {author} {\bibfnamefont {G.~S.}\ \bibnamefont {Grest}}, \bibinfo
  {author} {\bibfnamefont {A.~Y.}\ \bibnamefont {Grosberg}}, \ and\ \bibinfo
  {author} {\bibfnamefont {K.}~\bibnamefont {Kremer}},\ }\href {\doibase
  10.1063/1.3587137} {\bibfield  {journal} {\bibinfo  {journal} {J. Chem.
  Phys.}\ }\textbf {\bibinfo {volume} {134}},\ \bibinfo {pages} {204904}
  (\bibinfo {year} {2011}{\natexlab{a}})}\BibitemShut {NoStop}%
\bibitem [{\citenamefont {Halverson}\ \emph
  {et~al.}(2011{\natexlab{b}})\citenamefont {Halverson}, \citenamefont {Lee},
  \citenamefont {Grest}, \citenamefont {Grosberg},\ and\ \citenamefont
  {Kremer}}]{Halverson2011a}%
  \BibitemOpen
  \bibfield  {author} {\bibinfo {author} {\bibfnamefont {J.~D.}\ \bibnamefont
  {Halverson}}, \bibinfo {author} {\bibfnamefont {W.~B.}\ \bibnamefont {Lee}},
  \bibinfo {author} {\bibfnamefont {G.~S.}\ \bibnamefont {Grest}}, \bibinfo
  {author} {\bibfnamefont {A.~Y.}\ \bibnamefont {Grosberg}}, \ and\ \bibinfo
  {author} {\bibfnamefont {K.}~\bibnamefont {Kremer}},\ }\href
  {https://aip.scitation.org/doi/pdf/10.1063/1.3587138} {\bibfield  {journal}
  {\bibinfo  {journal} {J. Chem. Phys.}\ }\textbf {\bibinfo {volume} {134}},\
  \bibinfo {pages} {204905} (\bibinfo {year} {2011}{\natexlab{b}})}\BibitemShut
  {NoStop}%
\bibitem [{\citenamefont {Everaers}\ \emph {et~al.}(2004)\citenamefont
  {Everaers}, \citenamefont {Sukumaran}, \citenamefont {Grest}, \citenamefont
  {Svaneborg}, \citenamefont {Sivasubramanian},\ and\ \citenamefont
  {Kremer}}]{Everaers2004}%
  \BibitemOpen
  \bibfield  {author} {\bibinfo {author} {\bibfnamefont {R.}~\bibnamefont
  {Everaers}}, \bibinfo {author} {\bibfnamefont {S.~K.}\ \bibnamefont
  {Sukumaran}}, \bibinfo {author} {\bibfnamefont {G.~S.}\ \bibnamefont
  {Grest}}, \bibinfo {author} {\bibfnamefont {C.}~\bibnamefont {Svaneborg}},
  \bibinfo {author} {\bibfnamefont {A.}~\bibnamefont {Sivasubramanian}}, \ and\
  \bibinfo {author} {\bibfnamefont {K.}~\bibnamefont {Kremer}},\ }\href@noop {}
  {\bibfield  {journal} {\bibinfo  {journal} {Science}\ }\textbf {\bibinfo
  {volume} {303}},\ \bibinfo {pages} {823} (\bibinfo {year}
  {2004})}\BibitemShut {NoStop}%
\bibitem [{\citenamefont {Ge}\ \emph {et~al.}(2014)\citenamefont {Ge},
  \citenamefont {Robbins}, \citenamefont {Perahia},\ and\ \citenamefont
  {Grest}}]{Ge2014}%
  \BibitemOpen
  \bibfield  {author} {\bibinfo {author} {\bibfnamefont {T.}~\bibnamefont
  {Ge}}, \bibinfo {author} {\bibfnamefont {M.~O.}\ \bibnamefont {Robbins}},
  \bibinfo {author} {\bibfnamefont {D.}~\bibnamefont {Perahia}}, \ and\
  \bibinfo {author} {\bibfnamefont {G.~S.}\ \bibnamefont {Grest}},\ }\href@noop
  {} {\bibfield  {journal} {\bibinfo  {journal} {Phys. Rev. E}\ }\textbf
  {\bibinfo {volume} {90}},\ \bibinfo {pages} {012602} (\bibinfo {year}
  {2014})}\BibitemShut {NoStop}%
\bibitem [{\citenamefont {Stevens}(2001)}]{Stevens2001}%
  \BibitemOpen
  \bibfield  {author} {\bibinfo {author} {\bibfnamefont {M.~J.}\ \bibnamefont
  {Stevens}},\ }\href@noop {} {\bibfield  {journal} {\bibinfo  {journal}
  {Macromolecules}\ }\textbf {\bibinfo {volume} {34}},\ \bibinfo {pages} {2710}
  (\bibinfo {year} {2001})}\BibitemShut {NoStop}%
\bibitem [{\citenamefont {Rottler}\ \emph {et~al.}(2002)\citenamefont
  {Rottler}, \citenamefont {Barsky},\ and\ \citenamefont
  {Robbins}}]{Rottler2002}%
  \BibitemOpen
  \bibfield  {author} {\bibinfo {author} {\bibfnamefont {J.}~\bibnamefont
  {Rottler}}, \bibinfo {author} {\bibfnamefont {S.}~\bibnamefont {Barsky}}, \
  and\ \bibinfo {author} {\bibfnamefont {M.~O.}\ \bibnamefont {Robbins}},\
  }\href {\doibase 10.1103/PhysRevLett.89.148304} {\bibfield  {journal}
  {\bibinfo  {journal} {Phys. Rev. Lett.}\ }\textbf {\bibinfo {volume} {89}},\
  \bibinfo {pages} {148304} (\bibinfo {year} {2002})}\BibitemShut {NoStop}%
\bibitem [{\citenamefont {Ge}\ \emph {et~al.}(2013)\citenamefont {Ge},
  \citenamefont {Pierce}, \citenamefont {Perahia}, \citenamefont {Grest},\ and\
  \citenamefont {Robbins}}]{Ge2013}%
  \BibitemOpen
  \bibfield  {author} {\bibinfo {author} {\bibfnamefont {T.}~\bibnamefont
  {Ge}}, \bibinfo {author} {\bibfnamefont {F.}~\bibnamefont {Pierce}}, \bibinfo
  {author} {\bibfnamefont {D.}~\bibnamefont {Perahia}}, \bibinfo {author}
  {\bibfnamefont {G.~S.}\ \bibnamefont {Grest}}, \ and\ \bibinfo {author}
  {\bibfnamefont {M.~O.}\ \bibnamefont {Robbins}},\ }\href {\doibase
  10.1103/PhysRevLett.110.098301} {\bibfield  {journal} {\bibinfo  {journal}
  {Phys. Rev. Lett.}\ }\textbf {\bibinfo {volume} {110}},\ \bibinfo {pages}
  {098301} (\bibinfo {year} {2013})}\BibitemShut {NoStop}%
\bibitem [{\citenamefont {Wang}\ and\ \citenamefont {Ge}(2021)}]{Wang2021}%
  \BibitemOpen
  \bibfield  {author} {\bibinfo {author} {\bibfnamefont {J.}~\bibnamefont
  {Wang}}\ and\ \bibinfo {author} {\bibfnamefont {T.}~\bibnamefont {Ge}},\
  }\href {\doibase 10.1021/acs.macromol.1c01080} {\bibfield  {journal}
  {\bibinfo  {journal} {Macromolecules}\ }\textbf {\bibinfo {volume} {54}},\
  \bibinfo {pages} {7500} (\bibinfo {year} {2021})}\BibitemShut {NoStop}%
\bibitem [{\citenamefont {Zhang}\ \emph {et~al.}(2011)\citenamefont {Zhang},
  \citenamefont {Lackey}, \citenamefont {Cui},\ and\ \citenamefont
  {Tew}}]{Zhang2011}%
  \BibitemOpen
  \bibfield  {author} {\bibinfo {author} {\bibfnamefont {K.}~\bibnamefont
  {Zhang}}, \bibinfo {author} {\bibfnamefont {M.~A.}\ \bibnamefont {Lackey}},
  \bibinfo {author} {\bibfnamefont {J.}~\bibnamefont {Cui}}, \ and\ \bibinfo
  {author} {\bibfnamefont {G.~N.}\ \bibnamefont {Tew}},\ }\href {\doibase
  10.1021/ja111391z} {\bibfield  {journal} {\bibinfo  {journal} {J. Am. Chem.
  Soc.}\ }\textbf {\bibinfo {volume} {133}},\ \bibinfo {pages} {4140} (\bibinfo
  {year} {2011})}\BibitemShut {NoStop}%
\bibitem [{\citenamefont {Trachsel}\ \emph {et~al.}(2021)\citenamefont
  {Trachsel}, \citenamefont {Romio}, \citenamefont {Zenobi‐Wong},\ and\
  \citenamefont {Benetti}}]{Trachsel2021}%
  \BibitemOpen
  \bibfield  {author} {\bibinfo {author} {\bibfnamefont {L.}~\bibnamefont
  {Trachsel}}, \bibinfo {author} {\bibfnamefont {M.}~\bibnamefont {Romio}},
  \bibinfo {author} {\bibfnamefont {M.}~\bibnamefont {Zenobi‐Wong}}, \ and\
  \bibinfo {author} {\bibfnamefont {E.~M.}\ \bibnamefont {Benetti}},\
  }\href@noop {} {\bibfield  {journal} {\bibinfo  {journal} {Macromol. Rapid
  Commun.}\ }\textbf {\bibinfo {volume} {42}},\ \bibinfo {pages} {2000658}
  (\bibinfo {year} {2021})}\BibitemShut {NoStop}%
\bibitem [{\citenamefont {Duering}\ \emph {et~al.}(1994)\citenamefont
  {Duering}, \citenamefont {Kremer},\ and\ \citenamefont
  {Grest}}]{Duering1994}%
  \BibitemOpen
  \bibfield  {author} {\bibinfo {author} {\bibfnamefont {E.~R.}\ \bibnamefont
  {Duering}}, \bibinfo {author} {\bibfnamefont {K.}~\bibnamefont {Kremer}}, \
  and\ \bibinfo {author} {\bibfnamefont {G.~S.}\ \bibnamefont {Grest}},\
  }\href@noop {} {\bibfield  {journal} {\bibinfo  {journal} {J. Chem. Phys.}\
  }\textbf {\bibinfo {volume} {101}},\ \bibinfo {pages} {8169} (\bibinfo {year}
  {1994})}\BibitemShut {NoStop}%
\bibitem [{\citenamefont {Rubinstein}\ and\ \citenamefont
  {Colby}(2003)}]{Rubinstein2003}%
  \BibitemOpen
  \bibfield  {author} {\bibinfo {author} {\bibfnamefont {M.}~\bibnamefont
  {Rubinstein}}\ and\ \bibinfo {author} {\bibfnamefont {R.~H.}\ \bibnamefont
  {Colby}},\ }\href@noop {} {\emph {\bibinfo {title} {Polymer Physics}}}\
  (\bibinfo  {publisher} {Oxford University Press},\ \bibinfo {year}
  {2003})\BibitemShut {NoStop}%
\bibitem [{not()}]{note1}%
  \BibitemOpen
  \bibinfo {note} {A recent simulation \cite{Wang2021} has shown that in the
  glassy state of ring polymers, which is far from equilibrium, a fraction of
  entanglements form a network that support the stable growth of craze
  fibrils.}\BibitemShut {Stop}%
\bibitem [{\citenamefont {Romio}\ \emph {et~al.}(2020)\citenamefont {Romio},
  \citenamefont {Trachsel}, \citenamefont {Morgese}, \citenamefont
  {Ramakrishna}, \citenamefont {Spencer},\ and\ \citenamefont
  {Benetti}}]{Romio2020}%
  \BibitemOpen
  \bibfield  {author} {\bibinfo {author} {\bibfnamefont {M.}~\bibnamefont
  {Romio}}, \bibinfo {author} {\bibfnamefont {L.}~\bibnamefont {Trachsel}},
  \bibinfo {author} {\bibfnamefont {G.}~\bibnamefont {Morgese}}, \bibinfo
  {author} {\bibfnamefont {S.~N.}\ \bibnamefont {Ramakrishna}}, \bibinfo
  {author} {\bibfnamefont {N.~D.}\ \bibnamefont {Spencer}}, \ and\ \bibinfo
  {author} {\bibfnamefont {E.~M.}\ \bibnamefont {Benetti}},\ }\href@noop {}
  {\bibfield  {journal} {\bibinfo  {journal} {ACS Macro Lett.}\ }\textbf
  {\bibinfo {volume} {9}},\ \bibinfo {pages} {1024} (\bibinfo {year}
  {2020})}\BibitemShut {NoStop}%
\bibitem [{\citenamefont {Xu}\ \emph {et~al.}(2018)\citenamefont {Xu},
  \citenamefont {Chen}, \citenamefont {Wang}, \citenamefont {Zheng},
  \citenamefont {Ding}, \citenamefont {Jiang}, \citenamefont {Tan},\ and\
  \citenamefont {Fu}}]{Fu2018}%
  \BibitemOpen
  \bibfield  {author} {\bibinfo {author} {\bibfnamefont {J.}~\bibnamefont
  {Xu}}, \bibinfo {author} {\bibfnamefont {W.}~\bibnamefont {Chen}}, \bibinfo
  {author} {\bibfnamefont {C.}~\bibnamefont {Wang}}, \bibinfo {author}
  {\bibfnamefont {M.}~\bibnamefont {Zheng}}, \bibinfo {author} {\bibfnamefont
  {C.}~\bibnamefont {Ding}}, \bibinfo {author} {\bibfnamefont {W.}~\bibnamefont
  {Jiang}}, \bibinfo {author} {\bibfnamefont {L.}~\bibnamefont {Tan}}, \ and\
  \bibinfo {author} {\bibfnamefont {J.}~\bibnamefont {Fu}},\ }\href@noop {}
  {\bibfield  {journal} {\bibinfo  {journal} {Chem. Mater.}\ }\textbf {\bibinfo
  {volume} {30}},\ \bibinfo {pages} {6026} (\bibinfo {year}
  {2018})}\BibitemShut {NoStop}%
\bibitem [{\citenamefont {Guo}\ \emph {et~al.}(2020)\citenamefont {Guo},
  \citenamefont {Han}, \citenamefont {Zhao}, \citenamefont {Yang},\ and\
  \citenamefont {Zhang}}]{Guo2020}%
  \BibitemOpen
  \bibfield  {author} {\bibinfo {author} {\bibfnamefont {H.}~\bibnamefont
  {Guo}}, \bibinfo {author} {\bibfnamefont {Y.}~\bibnamefont {Han}}, \bibinfo
  {author} {\bibfnamefont {W.}~\bibnamefont {Zhao}}, \bibinfo {author}
  {\bibfnamefont {J.}~\bibnamefont {Yang}}, \ and\ \bibinfo {author}
  {\bibfnamefont {L.}~\bibnamefont {Zhang}},\ }\href {\doibase
  10.1038/s41467-020-15949-8} {\bibfield  {journal} {\bibinfo  {journal} {Nat.
  Commun.}\ }\textbf {\bibinfo {volume} {11}},\ \bibinfo {pages} {2037}
  (\bibinfo {year} {2020})}\BibitemShut {NoStop}%
\bibitem [{\citenamefont {Sun}\ \emph {et~al.}(2012)\citenamefont {Sun},
  \citenamefont {Zhao}, \citenamefont {Illeperuma}, \citenamefont {Chaudhuri},
  \citenamefont {Oh}, \citenamefont {Mooney}, \citenamefont {Vlassak},\ and\
  \citenamefont {Suo}}]{Sun2012}%
  \BibitemOpen
  \bibfield  {author} {\bibinfo {author} {\bibfnamefont {J.-Y.}\ \bibnamefont
  {Sun}}, \bibinfo {author} {\bibfnamefont {X.}~\bibnamefont {Zhao}}, \bibinfo
  {author} {\bibfnamefont {W.~R.}\ \bibnamefont {Illeperuma}}, \bibinfo
  {author} {\bibfnamefont {O.}~\bibnamefont {Chaudhuri}}, \bibinfo {author}
  {\bibfnamefont {K.~H.}\ \bibnamefont {Oh}}, \bibinfo {author} {\bibfnamefont
  {D.~J.}\ \bibnamefont {Mooney}}, \bibinfo {author} {\bibfnamefont {J.~J.}\
  \bibnamefont {Vlassak}}, \ and\ \bibinfo {author} {\bibfnamefont
  {Z.}~\bibnamefont {Suo}},\ }\href
  {https://www.nature.com/articles/nature11409.pdf} {\bibfield  {journal}
  {\bibinfo  {journal} {Nature}\ }\textbf {\bibinfo {volume} {489}},\ \bibinfo
  {pages} {133} (\bibinfo {year} {2012})}\BibitemShut {NoStop}%
\end{thebibliography}



\end{document}